\documentclass[aps,pre,showpacs,twocolumn]{revtex4}

\usepackage{graphics}
\usepackage{epsfig}
\begin{document}
\title{Non-Rayleigh limit of the Lorenz-Mie solution and suppression of the scattering\\ by the spheres of negative refractive index}

\author{Andrey E. Miroshnichenko}
\email{aem124@rsphysse.anu.edu.au}
\affiliation{Nonlinear Physics Centre and Centre for Ultra-high bandwidth Devices for Optical Systems (CUDOS), Australian National University, Canberra ACT 0200, Australia}

\begin{abstract}
We study light scattering by spheres made of negative refractive index materials. We demonstrate that scattering efficiency of the light does not always obey the well-known Rayleigh's law $Q_{\rm sca}\sim1/\lambda^4$ for small values of size parameter, but could be wavelength independent $Q_{\rm sca}\sim const$, or even increase  with the wavelength as $Q_{\rm sca}\sim\lambda^2$. In addition, it is found that spheres with radius comparable with the input wavelength may exhibit suppression of the scattering in any directions, particularly in forward one. It can be considered as analogues  of Brewster's angles for spheres.
\end{abstract}

\pacs{42.25.Bs, 42.68.Mj, 92.60.Ta}

\maketitle

\section{Introduction}
Light scattering by small particles is one of the fundamental problems, which found a broad application in various fields such as meteorology, biology and medicine. The first results on scattering by small particles were derived by Lord Rayleigh~\cite{lr:pm:71,lr:pm:71b,lr:pm:71c}, and analyzed by J.C. Maxwell-Garnett~\cite{jcmg:ptrslsa:04,jcmg:ptrslsa:06}. It is widely excepted that the complete solution for the homogeneous spheres of arbitrary size and optical constants was found by Mie more than a century ago~\cite{gm:ap:08}.  As a matter of fact, the first exact solution was obtained by Lorenz in 1890~\cite{lvl:kdvss:90}, based on his own version of the electromagnetism. But, this paper was published in Danish, and did not receive a proper acknowledgement by the scientific community~\cite{hk:ao:91}. Thus, it will be reasonable from a historical point of view to call it Lorenz-Mie solution. This exact solution was carefully analyzed, and successfully applied to describe scattering by particles of various shapes. In the limit of small spheres it reproduces the results obtained by Lord Rayleigh. However, recent studies of resonant scattering by small particles with weak dissipation rates~\cite{mbvfnz:oe:05,mitbsl:prl:06} have revealed novel and unexpected features, namely giant optical resonances with an inverse hierarchy (the quadrupole resonance is much stronger than the dipole one, etc.)~\cite{mit:spj:84} , a complicated near-field structure with vortices~\cite{bslmitvtzbwmhhlptcc:jopao:07}, unusual frequency and size dependencies~\cite{lbstmiwzbzyhmhlpctc:apmsp:07}, which allow to name such a scattering {anomalous}.

Nowadays technological success in fabrication of nanostructured materials has led to 
an emergence of the rapidly growing field of nanophotonics. The primary interest of this field is to achieve an enhancement of light-matter interaction at the nanoscale. The potential for this is based on the possibility to excite  localized surface plasmons in metallic nanoparticles. The strong light confinement in ultra-small volumes may result in various nonlinear effects. The Lorenz-Mie solution is able to shed a light on the origin of such enhancement by metal nanoparticles, and to provide better understanding of the phenomenon.

The beginning of XXI century was marked by opening a new chapter of electrodyanical theory, when it was generally accepted by scientific community that material with negative refractive index can exist, and can even be fabricated. A lot of attention was attracted to such kind of materials with negative index of refraction since they possess unusual optical properties such as inverted Snell's law~\cite{vgv:spu:68}, backward waves propagation, possibility to design perfect lens~\cite{jbp:prl:00}, and even cloak of invisibility~\cite{jbpdsdrs:s:06}. The origin of such peculiar effects lies in the interplay  of dielectric and magnetic responses. Almost every corner of electrodynamic theory was revised from the perspective of materials with negative refractive index with aim to find novel phenomena, which were overlooked in the past due to restricted parameters range.

In this paper we study light scattering by spheres made of materials with negative refractive index. For that purpose we are using exact Lorenz-Mie solution, which is valid for any values of permettivities $\epsilon_s$ and permiabilities $\mu_s$, including negative ones, and various ratios $\kappa=2\pi a/\lambda$ of radius of the sphere $a$ to the wavelength $\lambda$. This solution predicts that spheres of the small radius $a\rightarrow0$ made of conventional materials will scatter light as $Q_{\rm sca}\sim1/\lambda^4$, which is in agreement with the Rayleigh scattering by small particles. It means that the blue component of the light is scattered 10 times stronger than the red one. We evidence this effect everyday by looking at the blue sky. Another prediction of Lorenz-Mie solution for conventional materials is that scattering by large spheres is predominantly in forward direction, which is knows as Mie scattering. According to our analysis of the Lorenz-Mie solution it turns out that both effects (Rayleigh and Mie scatterings) can be altered by negative refractive index materials. In particular, the light scattering by small spheres can be wavelength independent $Q_{\rm sca}\sim const$, or even diverge $Q_{\rm sca}\sim\lambda^2$. In addition to this, the scattering can be significantly suppressed in any direction, including the forward one.

The paper is organized as follows. In Section II the Lorenz-Mie solution is described. The scattering by small sphere with negative refractive index is studied in Sec. III, where different regimes are identified. In Sec. IV the angular dependence of scattered light is considered. Section V concludes the paper.

\section{Lorenz-Mie solution}
We consider the problem of light scattering by a spherical particle of radius $a$, made of an material with arbitrary permettivity $\epsilon_s$ and permiability $\mu_s$, embedded into surrounding media with given optical constants $\epsilon_o$ and $\mu_o$. For that purpose we will utilize the exact Lorenz-Mie solution~\cite{gm:ap:08,bh}. According to this solution the scattering efficiency of light can be written as follows
\begin{eqnarray}\label{eq:eq1}
Q_{\rm sca} = \frac{2}{x^2}\sum\limits_{n=1}^{\infty}(2n+1)(|a_n|^2+|b_n|^2)\;,
\end{eqnarray}
where $\kappa=2\pi a/\lambda$ is the size parameter, $a_n$ and $b_n$ are scattering coefficients. The latter ones can be expressed in terms of Riccati-Bessel functions of the first and third kind
\begin{eqnarray}\label{eq:eq2}
a_n &=& \frac{\eta\psi_n(m\kappa)\psi_n^{\prime}(\kappa)-\psi_n(\kappa)\psi^{\prime}(m\kappa)}{\eta\psi_n(m\kappa)\xi_n^{\prime}(\kappa)-\xi_n(\kappa)\psi_n^{\prime}(m\kappa)}\;,
\end{eqnarray}
where $m=\sqrt{\epsilon_s\mu_s/\epsilon_o\mu_o}$ is the ratio of refractive indices, and $\eta=m\mu_o/\mu_s=\sqrt{\epsilon_s\mu_o/\epsilon_o\mu_s}$ is the ratio of electromagnetic impedances of the sphere and surrounding media. Due to the symmetry of the scattering coefficients~\cite{gvwsb:pr:92} the expression for $b_n$ coefficients can be obtained as follows
\begin{eqnarray}\label{eq:eq3}
b_n(\eta,m,\kappa)=a_n(1/\eta,m,\kappa)\;.
\end{eqnarray}
In the case of surrounding media with equal optical constants $\epsilon_o=\mu_o$ (air, for example), there are other symmetries of the scattering coefficients, which will be used later. When the permettivity and pemiability of the sphere are the same $\epsilon_s=\mu_s$, the parameter $\eta$ reduces to unity. Due to relation (\ref{eq:eq3}) the scattering coefficients become also identical $a_n=b_n$. The substitution  $\epsilon_s\rightarrow\tilde{\mu}_s$ and $\mu_s\rightarrow\tilde{\epsilon}_s$ results in $\eta\rightarrow1/\tilde{\eta}$, which again due to relation (\ref{eq:eq3}) will simply swap scattering coefficients $a_n\rightarrow\tilde{b}_n$ and $b_n\rightarrow\tilde{a}_n$. Both symmetries leave the scattering efficiency (\ref{eq:eq1}) unchanged. Because of these symmetries, it will be enough to consider materials, whose optical properties satisfy the relation $|\epsilon_s|\le|\mu_s|$.

In the case of nonabsorbing sphere, $Im(\epsilon_s)=Im(\mu_s)=0$, the scattering efficiency is equal to the extinction efficiency $Q_{\rm sca}=Q_{\rm ext}$, and, thus, describes the overall attenuation of the light as it propagates through the sphere.

The angular dependence of the scattered light can be studied by using scattering amplitudes
\begin{eqnarray}\label{eq:eq4}
S_{\bot}(\theta)&=&\sum\limits_{n=1}^{\infty}\frac{2n+1}{n(n+1)}(a_n\pi_n+b_n\tau_n)\;,\\
S_{||}(\theta)&=&\sum\limits_{n=1}^{\infty}\frac{2n+1}{n(n+1)}(a_n\tau_n+b_n\pi_n)\nonumber
\end{eqnarray}
of perpendicular $S_{\bot}$ and parallel $S_{||}$ polarized light to the scattered plane, determined by the incident and scattered directions. The functions $\pi_n$ and $\tau_n$ can be defined via Legendre polynomials
\begin{eqnarray}\label{eq:ea5}
\pi_n = \frac{P_n}{\sin\theta}\;,\;\;\; \tau_n = \frac{dP_n}{d\theta}\;.
\end{eqnarray}
From expressions (\ref{eq:eq4}) we can define transmission $T$ (forward, $\theta=0^\circ$) and reflection $R$ (backward, $\theta=180^\circ$) amplitudes
\begin{eqnarray}\label{eq:eq6}
T& = &S_{\bot,||}(0^\circ)  = \frac{1}{2}\sum\limits_{n=1}^{\infty}(2n+1)(a_n+b_n)\;,\\
R&=&\pm S_{\bot,||}(180^\circ) = \frac{1}{2}\sum\limits_{n=1}^{\infty}(-1)^n(2n+1)(a_n-b_n)\;.\nonumber
\end{eqnarray}

These two directions are special ones, since the scattering does not depend on the input polarization.

\section{(non) Rayleigh limit}
For spheres with the radius much smaller than wavelength of the light $\kappa\ll1$  the general solution (\ref{eq:eq1}) converges very fast, and it is possible to retain only the first terms $a_1$ and $b_1$, since all other are negligible $|a_n|\approx|b_n|\ll0$ for $n>1$. 
The Riccati-Bessel functions for $n=1$ have a very simple form~\cite{bh}
\begin{eqnarray}\label{eq:eq7}
\psi_1(\rho)=\frac{\sin(\rho)}{\rho}-\cos(\rho)\;,\;\;\xi_1(\rho)=-\exp(i\rho)(\frac{i}{\rho}+1)\;.
\end{eqnarray}
To make our analysis applicable to any values of optical constants $\epsilon_s$ and $\mu_s$ we will use relations (\ref{eq:eq7}) for functions with argument $m\kappa$, and their expansions up to the second order~\cite{gvwsb:pr:92}
\begin{eqnarray}\label{eq:eq8}
\psi_1(\kappa)\sim\frac{\kappa^2}{3}\;,\;\;\xi_1(\kappa)\sim-\frac{i}{\kappa}-\frac{i\kappa}{2}+\frac{\kappa^2}{3}
\end{eqnarray}
for functions with argument $\kappa$ in expressions (\ref{eq:eq2}).
Thus, after some algebra the scattering coefficient $a_1$ can be written in the following form

\begin{widetext}
\begin{eqnarray}\label{eq:eq9}
a_1=\frac{A1}{A2}=\frac{2\kappa^3\{[1+2\eta m][mx\cos(m\kappa)-\sin(m\kappa)]+m^2\kappa^2\sin(m\kappa)\}}{\eta m[6i-3i\kappa^2+4\kappa^3][m\kappa\cos(m\kappa)-\sin(m\kappa)]+[2\kappa^3-6i-3i\kappa^2][m\kappa\cos(m\kappa)-(1-m^2\kappa^2)\sin(m\kappa)]}\;.
\end{eqnarray}
\end{widetext}
The expression for $b_1$ scattering coefficient can be obtained by using the symmetry relation (\ref{eq:eq3}).

To make a proper analysis of the scattering coefficient $a_1$ in the limit of the small size parameter $\kappa\ll1$, we expand the numerator $A1$ and denominator $A2$ separately up to the same order
\begin{eqnarray}\label{eq:eq10}
A1&=&-\frac{4}{3}(\eta m-1)m^3\kappa^6+O(\kappa^7)\;,\nonumber\\
A2&=&-2i(\eta m+2)m^3\kappa^3+\ldots\\
&&+\frac{i}{5}[4m^2-10+\eta m(5+m^2)]m^3\kappa^5-\ldots\nonumber\\
&&-\frac{4}{3}(\eta m-1)m^3\kappa^6+O(\kappa^7)\;.\nonumber
\end{eqnarray}

In general only leading orders are kept in $A1$ and $A2$, which allow one to simplify the scattering coefficients $a_1$ and $b_1$ to the form~\cite{mkdwclg:josa:83}
\begin{eqnarray}\label{eq:eq11}
a_1 &\approx& -i\frac{2}{3}\frac{\eta m-1}{\eta m+2}\kappa^3= -i\frac{2}{3}\frac{\epsilon_s-\epsilon_o}{\epsilon_s+2\epsilon_o}\kappa^3\;,\\
b_1 &\approx& -i\frac{2}{3}\frac{m-\eta}{m+2\eta}\kappa^3 = -i\frac{2}{3}\frac{\mu_s-\mu_o}{\mu_s+2\mu_o}\kappa^3\;.\nonumber
\end{eqnarray}
The substitution of these relations into the expression of the scattering efficiency (\ref{eq:eq1}) will result in
\begin{eqnarray}\label{eq:eq12}
Q_{\rm sca} \approx \frac{8}{3}\left[\left|\frac{\epsilon_s-\epsilon_o}{\epsilon_s+2\epsilon_o}\right|^2+\left|\frac{\mu_s-\mu_o}{\mu_s+2\mu_o}\right|^2\right]\kappa^4\;,
\end{eqnarray}
which is in agreement with Rayleigh's approach, according to which small spheres effectively scatter light as $Q_{\rm sca}\sim1/\lambda^4$. This is a well-known result which is applicable to spheres made of almost any materials, except for $\epsilon_s = -2\epsilon_o$ or $\mu_s = -2\mu_o$. In this case one of the scattering coefficient diverges (\ref{eq:eq11}) , and so does the scattering efficiency (\ref{eq:eq12}). Usually, this divergence is associated with the excitation of a localized surface modes of spheres of the larger radius $\kappa>1$~\cite{bh}. But, this interpretation is not applicable in the limit of small spheres $\kappa\ll1$, and should be revised.

\begin{figure}
\vspace{20pt}
\includegraphics[width=1.\columnwidth]{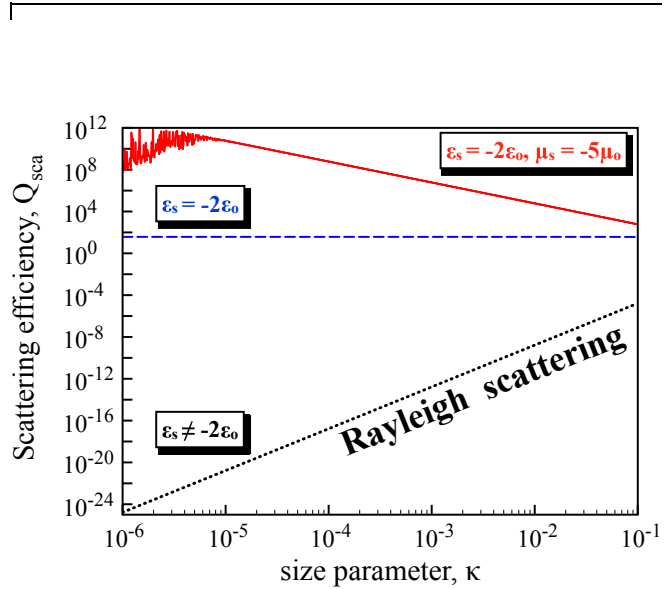}%
\caption{\label{fig:fig1} (Color online)
Loglog plot of the scattering efficiency dependence vs size parameter $\kappa$, calculated by using the exact Lorenz-Mie solutions for the sphere embedded into air ($\epsilon_o=\mu_o=1$). We chose  three sets of parameters to illustrate different types of the asymptotic behavour : i) $\epsilon_s=2$ and $\mu_s=1$ (dotted line); ii) $\epsilon_s = -2$ and $\mu_s = -3$ (dashed line), iii) $\epsilon_s=-2$ and $\mu_s=-5$ (solid line). The calculated results are in very good agreement with the analytical ones (\ref{eq:eq12},\ref{eq:eq14},\ref{eq:eq16}).
Numerical errors in calculations of the Bessel functions become essential for the very small size parameter $\kappa<10^{-5}$.
}
\end{figure}

Indeed, the relations (\ref{eq:eq11}) are only expansions of the scattering coefficients $a_1$ and $b_1$ (\ref{eq:eq9}). And if an expansion produces some divergences we should look back into the original expression and take a special care for these particular parameters. In the expressions (\ref{eq:eq10}) the first term of the denominator $A2$ vanishes exactly when $\epsilon_s = -2\epsilon_o$. It means that we should take the second term to obtain the correct expression for the scattering coefficient $a_1$
\begin{eqnarray}\label{eq:eq13}
a_1 \approx \frac{5i\mu_o}{\mu_s+5\mu_o}\kappa\;,
\end{eqnarray}
while the leading order of $b_1$ scattering coefficient remains unchanged.
The scattering efficiency in this case becomes {\em wavelength independent}\begin{eqnarray}\label{eq:eq14}
Q_{\rm sca} \approx 150\left|\frac{\mu_o}{\mu_s+5\mu_o}\right|^2\;.
\end{eqnarray}

One may notice that if in addition to the condition $\epsilon_s = -2\epsilon_o$ the permeability of the sphere satisfies the relation $\mu_s = -5\mu_o$ both the scattering coefficient $a_1$ (\ref{eq:eq13}) and efficiency $Q_{\rm sca}$ (\ref{eq:eq14}) will diverge.
It means that one should revise the expression for the denominator $A2$ again. For these particular parameters first two terms vanish in $A2$  (\ref{eq:eq10}), and the third term is identical to the nominator $A1$. It results in the following scattering coefficients
\begin{eqnarray}\label{eq:eq15}
a_1\approx 1\;,\;\;\; b_1\approx-\frac{4i}{3}\kappa^3\;.
\end{eqnarray}
The scattering efficiency in this case can be written as
\begin{eqnarray}\label{eq:eq16}
Q_{\rm sca}\approx\frac{6}{\kappa^2}\;,
\end{eqnarray}
which is proportional to the square of the  wavelength $Q_{\rm sca}\sim\lambda^2$.
We would like to note here, that the expression similar to eq. (\ref{eq:eq9}) was obtained in Ref.~\cite{bgfmfgjmsgv:josa:08} without subsequent analysis. We found it incorrect since it does not reproduce the Rayleigh scattering regime, although it still allows to predict correctly other limits.

The scattering efficiencies calculated by using the exact Lorenz-Mie solution are shown in Fig.\ref{fig:fig1}. The asymptotic  behaviour of all curves are in a very good agreement with analytical derivations (\ref{eq:eq12},\ref{eq:eq14},\ref{eq:eq16}), except the very small values of the size parameter $x<10^{-5}$ because of the difficulties of numerical calculations of the Bessel functions~\cite{bh}. 

\begin{figure}
\vspace{20pt}
\includegraphics[width=1.\columnwidth]{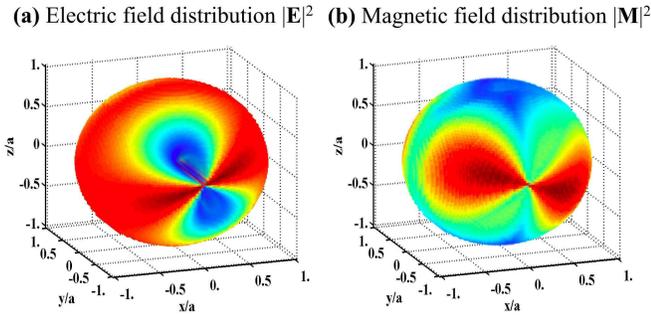}%
\caption{\label{fig:fig2} (Color online)
The distribution of the electric (a) and magnetic (b) fields inside the sphere of radius $a=10^{-5}$.
}
\end{figure}

Naturally, the question arises about the origin of changes of the scattering efficiency asymptotic behaviour. As it was mentioned above, the argument of localized surface mode excitation is not applicable here, because of the extremely small radius of the sphere compared to the wavelength. To answer this question, we calculate the distribution of the electric and magnetic fileds inside the sphere. It turns out that the distribution of the electric field is the same for all parameters, and resembles the electric dipole excitation [see Fig.\ref{fig:fig2}(a)], while the intensity scales with actual values of the optical constants. But what is changing inside the sphere is the distribution of the magnetic field. When $\epsilon_s\not=-2\epsilon_o$ the magnetic field inside the sphere is almost homogeneous and practically zero. The drastic change of the magnetic field distribution takes place when the condition $\epsilon_s=-2\epsilon_o$ is satisfied. It is no longer homogeneous, and resembles magnetic dipole excitation [see Fig.\ref{fig:fig2}(b)]. In general, under this condition, the magnetic field is still very weak, but becomes order of unity when the additional condition $\mu_s=-5\mu_o$ is satisfied. Thus, it allows us to conclude that the excitation of the magnetic dipole is responsible  for the variation of the asymptotic behaviour of the scattering efficiency. 

\section{Suppression of the scattering efficiency}
The analysis of the angular dependence of the scattered light (\ref{eq:eq4}) suggests that the scattering in the backward direction  (\ref{eq:eq6}) can be totally suppressed $R=0$ when the scattering coefficients are the same $a_n=b_n$ for all $n$. This condition is realized when the optical properties of the sphere are identical $\epsilon_s=\mu_s$ (\ref{eq:eq2}-\ref{eq:eq3}), due to the symmetry of the electric and magnetic fields. On contraray, the scattering in the forward direction  (\ref{eq:eq6}) can be suppressed $T=0$, when the scattering coefficients are opposite $a_n=-b_n$ for all $n$. Note here, that in both cases the strength of the  overall scattering efficiency is the same (\ref{eq:eq1}), indicating the fact that the sphere is equally excited under both conditions.  Obviously, they can not be exactly satisfied simultaneously, otherwise  $a_n=b_n\equiv0$.

As it was shown by Kerker~\cite{mkdwclg:josa:83} suppression in the forward direction can be achieved for small magnetic spheres. Indeed,  according to (\ref{eq:eq11}) for magnetic spheres embeded into air the condition $a_1=~-~b_1$ is fulfilled  for
\begin{eqnarray}\label{eq:eq17}
\epsilon_s=\frac{4-\mu_s}{1+2\mu_s}\;.
\end{eqnarray}
In Ref.~\cite{bgfgfmjms:josa:08} it was demonstrated that there is one exception of this formula, namely $\epsilon_s=\mu_s=-2$, which does not produce zero scattering in the forward direction. 
This is a very interesting case, because at the first sight it appears that both conditions for zero scattering  in forward $a_1=-b_1$ [because of (\ref{eq:eq17})] and backward $a_1=b_1$ [because of $\epsilon_s=\mu_s$] directions are satisfied, which should require that both scattering coefficient should vanish $a_1=b_1\equiv0$, suggesting that sphere is transparent, which is not true~\cite{bgfgfmjms:josa:08}. But the answer to this paradox is quite simple.
Our analysis in the previous section  of the scattering coefficients for the very small spheres suggests that expansions  (\ref{eq:eq11}) are no longer applicable because of divergence at $\epsilon_s=-2$, and the proper expansions (\ref{eq:eq13}) should be used instead, according to which $a_1\not=-b_1$. As a result,  the relation (\ref{eq:eq17}) is not applicable in this case, and, therefore, the scattering in the forward direction is nonzero.

\begin{figure}
\vspace{20pt}
\includegraphics[width=1.\columnwidth]{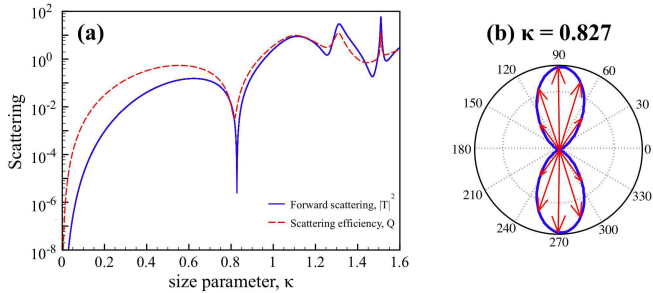}%
\caption{\label{fig:fig3} (Color online)
The scattering by the sphere with optical constants $\epsilon_s=\mu_s=-5$: (a) the scattering efficiency $Q_{\rm sca}$ (dashed line) and the forward scattering $|T|^2$ (solid line) vs size parameter $\kappa$, (b) angular dependence of the light scattering in the vicinity of the suppression of the forward scattering at $\kappa\approx0.827$. In this case the scattering is predominantly in the transverse directions.
}
\end{figure}

It is known that for spheres of large radius the scattering occurs predominantly in the forward direction $T\gg S_{\bot,||}(\theta\not=0^\circ)$~\cite{bh}. But, recently we has demonstrated~\cite{mitsfaemavgysk:prl:08} that weakly absorbing particles can exhibit suppression of the scattering in the forward direction, which can be explained in terms of the Fano resonance. This analysis is quite general, and can be extended to magnetic spheres as well. 
In Figure~\ref{fig:fig3}(a) we plot the dependence of the forward scattering $T$ versus size parameter $\kappa$ for optical constants $\epsilon_s=\mu_s=-5$. It indicates that near the size parameter value $\kappa\approx0.827$ there is a resonant suppression of the forward scattering $T\rightarrow0$. According to the optical theorem~\cite{bh}, which can be written in the absorptionless case as
\begin{eqnarray}\label{eq:eq18}
Q_{\rm ext}=Q_{\rm sca}=\frac{\lambda^2}{\pi}Re(T)\;,
\end{eqnarray}
the suppression of the scattering in the forward direction will result in the overall suppression of the extinction of light, making the sphere effectively transparent. 
Indeed, Fig.~\ref{fig:fig3}(a) clearly demonstrates that there is suppression of the scattering efficiency in the vicinity of the suppression in the forward direction. It is interesting to look at the angular dependence of the scattered light, since in this particular case the condition for zero backward scattering for any value of the size parameter is satisfied $\epsilon_s=\mu_s$. The resulting plot is showing in Fig.~\ref{fig:fig3}(b). It resembles figure-8 profile. It means, that all light is mostly scattered in transverse directions only. This is quite unusual behaviour, and it is opposite to the Rayleigh angular scattering, which is 90$^\circ$ rotated, and resembles figure-$\infty$. 

This analysis can be extended to the scattering in any direction $0^\circ<\theta<180^\circ$. In general, each polarization $S_{\bot,||}$ will be scattered differently (\ref{eq:eq4}).
 In case of magnetic particles it is possible to find the conditions when one of the polarizations will be strongly suppressed, resulting in totally polarized scattered light. These conditions can be considered as generalized Brewster's angles of the sphere. In case of the sphere of the small radius, such conditions can be found analytically from (\ref{eq:eq4}) and (\ref{eq:eq6}) with $\pi_1=1$ and $\tau_1=\cos\theta$
\begin{eqnarray}\label{eq:eq19}
\cos\theta_{\bot}^*=\frac{1}{\cos\theta_{||}^*}=\frac{(\epsilon_s-\epsilon_o)(\mu_s+2\mu_o)}{(\epsilon_s+2\epsilon_o)(\mu_s-\mu_o)}\;,
\end{eqnarray}
according to which $S_{\bot}(\theta_{\bot}^*)=0$ and $S_{||}(\theta_{||}^*)=0$.
The complete analysis of the angular depended scattering by small magnetic spheres can be found in Ref.~\cite{mkdwclg:josa:83}.

\begin{figure}
\vspace{20pt}
\includegraphics[width=1.\columnwidth]{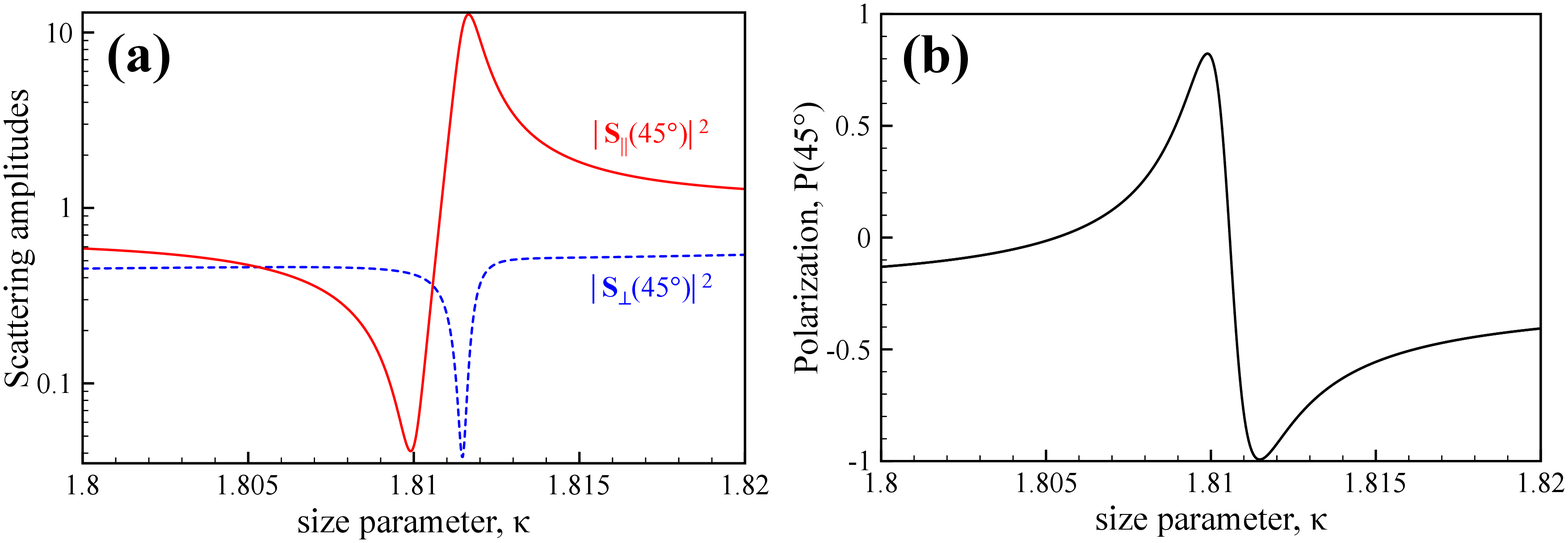}%
\caption{\label{fig:fig4} (Color online)
The scattering by the sphere with optical constants $\epsilon_s=-1$ and $\mu_s=-5$ in $\theta^*=45^\circ$ direction vs size parameter $\kappa$: (a) scattered perpendicular $S_{\bot}$ (dashed line) and parallel $S_{||}$ (solid line) polarized light; (b) total polarization $P$ according to eq.(\ref{eq:eq21}).  
}
\end{figure}

For spheres with the radius larger or comparable with the wavelength we can perform numerical analysis in any direction $\theta^*$, similar to the forward scattering in Fig.~\ref{fig:fig3}(a). The dependences of the scattering amplitudes (\ref{eq:eq4}) versus size parameter $\kappa$ for optical constants $\epsilon_s=-1$ and $\mu_s=-5$ in the direction $\theta^*=45^\circ$ are presented in Fig.~\ref{fig:fig4}(a). This particular example illustrates that when one of the scattering components $S_{\bot}$ is resonantly suppressed another one $S_{||}$ is resonantly enhanced in the vicinity of $\kappa\approx1.8115$. In addition, there is resonant suppression of $S_{||}$ at $\kappa\approx1.81$. All of these results in a sharp variation of the polarization
\begin{eqnarray}\label{eq:eq21}
P=\frac{|S_{\bot}|^2-|S_{||}|^2}{|S_{\bot}|^2+|S_{||}|^2}\;.
\end{eqnarray}
The scattered light is almost perpendicular polarized $|S_{\bot}|^2\approx1$ at $\kappa\approx1.81$,  and becomes totally parallel polarized $|S_{||}|^2=1$ at $\kappa\approx1.8115$ [see Fig.~\ref{fig:fig4}(b)]. In other words, by changing slightly the input wavelength, the scattered light in $\theta^*=45^\circ$ will swap it polarization. One might notice an asymmetric profile of resonant shapes, which, actually, can be understood in terms of the Fano resonance~\cite{mitsfaemavgysk:prl:08}.
 
\section{Conclusions}
We have studied light scattering by spheres made of materials with negative refractive index. By using the exact Lorenz-Mie solution we have demonstrated that in addition to the well-known wavelength dependence of the scattering efficiency $Q_{\rm sca}\sim1/\lambda^{-4}$, the negative index materials may inverse this dependence $Q_{\rm sca}\sim\lambda^{2}$, or even lead to the wavelength independent behaviour $Q_{\rm sca}\sim const$. The standard wavelength dependence was obtained by Lord Rayleigh in the assumption that small sphere can be represented by an electric dipole. Our analysis suggests that for certain values of the optical parameters $\epsilon_s$ and $\mu_s$ of the sphere a magnetic dipole may become excited as well. Its excitation results in a qualitatively different wavelength dependence of the scattering efficiency. Such regions have been found analytically, and confirmed numerically by using the Lorenz-Mie solution. We have also analyzed angular dependence of the scattered light for spheres with the radius comparable with the wavelength. It was found that both forward and backward scattering can be effectively suppressed leading the scattering predominantly in the transverse directions. Recent experimental observations~\cite{rvmrprvu:prcmmp:06,rvmrprdrvukp:prl:06,bgfmfgjms:prl:07,hrnk:prl:08} of the suppression of the forward scattering by magnetic spheres are in agreement with our results.
Moreover, we have demonstrated that the suppression of the scattering can be achieved in any given direction for various input polarization, providing with totally polarized scattered light. In analogous with the Brewster's angle at the interface of two media, such angles can be called Brewster's angles of the sphere.

It is worthwhile noting that the prefactor $(\epsilon_s-\epsilon_m)/(\epsilon_s+2\epsilon_m)$ in expression of the scattering coefficient $a_1$ (\ref{eq:eq11}) is exactly the same as Clausius-Mossotti factor in Lorenz-Lorentz formula, describing the relation between the refractive indices and polarizability. Our analysis suggests that in the case of negative refractive index materials, and in the limiting cases considered above, the Lorenz-Lorentz formula might be also changed according to eqs.(\ref{eq:eq13},\ref{eq:eq15}), which will require a separate study.

Thus, if the success in fabrication of metamaterials with negative refractive indices will be achieved we will be able, after all, to alter the familiar colour of the sky to be red, or even white.

\section*{Acknowledgments}
The author thanks Prof. Yuri Kivshar for useful discussions. 
The work has been supported by the Australian Research Council through the Discovery and Center of Excellence projects.

\bibliography{mie_negative_minimal}

\end{document}